%% file: paper.tex
\documentclass[aps,prl,showpacs,showkeys,twocolumn,superscriptaddress,
	      floats,floatfix,preprintnumbers,subfigure,nofootinbib]{revtex4-1}
\usepackage{amssymb,amsmath}
\usepackage{bm,url,graphics,subfigure,epsfig}
\usepackage{color}
\usepackage{tikz}

\input{symdef.inc}
\newcommand*\circled[1]{\tikz[baseline=(char.base)]{
            \node[shape=circle,draw,inner sep=0.5pt] (char) {#1};}}

\begin{document}

\title{Bifurcation dynamics of a particle-encapsulating droplet in shear flow}

\author{Lailai Zhu}
\email{lailaizhu00@gmail.com}
\affiliation{Laboratory of Fluid Mechanics and Instabilities, 
  Ecole Polytechnique F{\'e}d{\'e}rale de Lausanne, Lausanne, CH-1015, Switzerland.}
\affiliation{Department of Mechanical and Aerospace Engineering, Princeton
University, Princeton, NJ 08544, US.}
\affiliation{Linn\'{e} Flow Centre and Swedish e-Science Research Centre (SeRC), KTH 
Mechanics, Stockholm, SE-10044, Sweden.}
\author{Fran\c{c}ois Gallaire}
\affiliation{Laboratory of Fluid Mechanics and Instabilities, 
  Ecole Polytechnique F{\'e}d{\'e}rale de Lausanne, Lausanne, CH-1015, Switzerland.}

\begin{abstract}
To understand the behavior of composite fluid particles such as nucleated cells and double-emulsions
in flow, we study a finite-size 
particle encapsulated in a deforming droplet under shear flow as a model system.
In addition to its concentric particle-droplet configuration, we numerically explore 
other eccentric and time-periodic equilibrium solutions, which emerge spontaneously 
via supercritical 
pitchfork and Hopf bifurcations. We present the loci of these solutions 
around the codimenstion-two point. We adopt a dynamical system approach to 
model and characterize the
coupled behavior of the two bifurcations. By exploring the flow fields and hydrodynamic
forces in detail, we identify the role of hydrodynamic particle-droplet interaction which
gives rise to these bifurcations.

\end{abstract}
\preprint{}
\maketitle

Droplets, capsules and vesicles in flow often exhibit interestingly
rich dynamics even in the linear shear flow~\cite{stone1994dynamics, smith2004encapsulated, sibillo2006drop, skotheim2007red, omori2012reorientation, 
kraus1996fluid,misbah2006vacillating,noguchi2007swinging,deschamps2009phase}. Despite the substantial work on the dynamics
of these soft systems enclosing homogeneous fluids, limited effort
has been directed to studying their behavior when they include an internal
structure. However, such a configuration is common
 in nature and engineering applications: cells like
leukocytes, and megakaryocytes contain nucleus up to $50 \sim 80\%$ 
of themselves in volume~\cite{turgeon2005clinical}; double-emulsions playing an important role
in chemical and pharmaceutical engineering are featured with a core-shell
geometry~\cite{stone1990breakup,utada2005monodisperse,guzowski2013custom}
; droplet-based encapsulation
for high-throughput biological assays utilizes droplets as micro-chambers to compartment
cells for analysis at the single-cell level, where the cell size can be 
comparable to the droplet size
in certain applications
~\cite{he2005selective,chabert2008microfluidic,mazutis2013single}.

\begin{figure}[!h]
{        
    \centering    
    \hspace{-1em}\includegraphics[scale = 0.32] 
{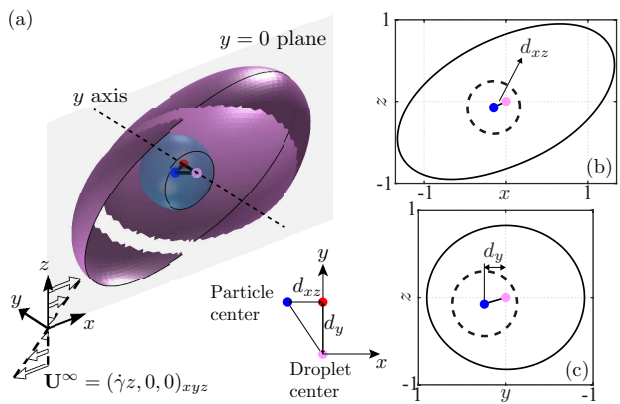}
\caption{(a) Sketch: a  spherical particle moving inside a droplet under shear. A 
snapshot of the composite system cut by the $y=0$ (b) and $x=0$ (c) 
plane.}
\label{fig:sketch}
}
\end{figure}

These fluid particles are characterized by complex hydrodynamic
interactions between the internal structures and the external 
interface. Few works conducted for
nucleated model cells in shear
~\cite{veerapaneni2011dynamics,kaoui2013complex,levant2014complex,luo2015deformation}
all assumed their compound structures to be concentric, 
preserving the rotational symmetry of order $2$ (C$2$) about the $y$ axis 
 and reflection symmetry about the $y=0$ shear plane (see Fig.~\ref{fig:sketch}a).
The symmetries do hold for a single shear-driven 
deformable particle which attains a steady ellipsoidal 
shape undergoing tank-treading motion~\cite{rallison1984deformation, seifert1997configurations, barthes2016motion}. Yet, they are 
not guaranteed in the presence of
an internal structure.

In this Letter we focus on the stability of the 
concentricity of composite fluid particles.
By considering a droplet encaging a spherical particle
as a model system, we formulate the following questions: will the 
composite structures remain concentric? How does the dynamics depend on
interfacial tension and particle size? 
What is the role of the hydrodynamic interaction?

We begin our discussion by presenting $3$D hydrodynamic simulations of 
a compound particle-droplet subjected to unbounded shear
$\bUinf = \bG  \cdot \bx $, in the creeping flow regime, where 
the only non-zero component  $G_{xz}=\shr$ represents the shear rate 
(Fig.~\ref{fig:sketch}a). The incompressible Stokes equations are solved by a boundary integral method 
(see Supplemental material and ~\cite{reigh2017swimming} for details). The immiscible Newtonian fluids
inside and outside the droplet have the same viscosity $\eta$; 
its surfactant-free interface has a uniform 
surface tension $\sigma$. 
The particle has a no-slip surface, freely translating and rotating
subject to zero hydrodynamic force and torque.
The droplet interface satisfies the standard stress balance 
condition~\cite{bc_drop,leal2007advanced}.
The radii of the particle and the undeformed droplet
are $a$ and $R$ respectively; the size ratio is denoted by $\alpha=a/R $ with $\alpha \in\lp 0, 1\rp$.
The capillary number $\Ca=\eta \shr R/\sigma$ indicates the 
ratio between viscous forces and capillary forces, 
limited to the regime without droplet break-up.
All length scales are scaled by $R$.

We initially displace the particle away
from the droplet center by a perturbative offset $\br=\lp \dx,\dy,\dz \rp$, then focusing on the time
evolution of $\dy$ and $\dxz = \sqrt{\dx^2+\dz^2}$ representing
the spanwise and in-plane displacements respectively. $\Dy$ and $\Dxz$ denote their equilibrium 
values when the system reaches a
steady or time-periodic state.

\begin{figure}[!ht]
{        
    \centering    
    \hspace{3em}\includegraphics[scale = 0.26] 
{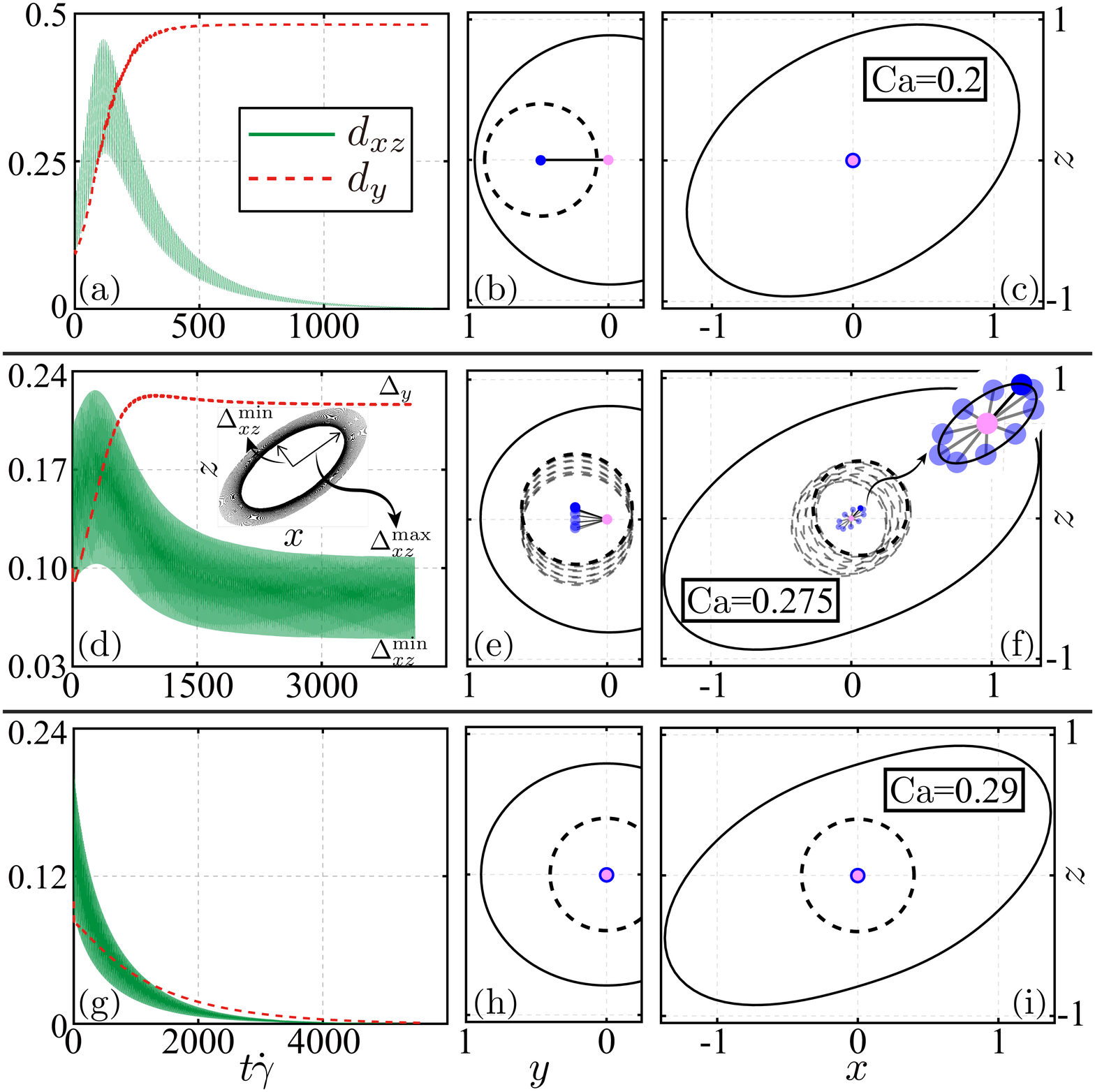}
\caption{Evolution of the in-plane and spanwise displacements $\dxz$ (green)
and $\dy$ (red) between the centers of particle (blue) and droplet (pink), for
size ratio $\alpha=0.4$ and $\Ca=0.2$ (first row), 
$0.275$ (second row) and $0.29$ (third row). The second (resp. third) column shows the profile of the composite
system at equilibrium, cut by $x=0$ (resp. $y=0$) plane.
The particle position is shown at different instants within a period for 
$\Ca=0.275$, the insets of (d) and (f) display the limit cycle solution.
}
\label{fig:d_v_time}
}
\end{figure}

We show the evolution of displacements in Fig.~\ref{fig:d_v_time}, 
presenting three typical $\Ca$-dependent scenarios for a 
particle of size ratio $\alpha=0.4$ (see Supplemental videos).
When $\Ca=0.2$, the in-plane 
displacement $\dxz$  decays asymptotically to zero after a transient growth while the spanwise offset 
$\dy$ increases to a saturated value $\Dy \approx 0.48$ indicating the broken reflection symmetry.
The particle rotates steadily near the lateral edge of the droplet interface (Fig.~\ref{fig:d_v_time}b). 
Increasing $\Ca$ to $0.275$, $\Dy$ decreases to $0.21$ approximately, while
$\dxz$ reaches a
time-periodic equilibrium cycle with a maximum of $\Dxzmax \approx 0.11$ and 
a minimum of $\Dxzmin \approx 0.04$. The particle follows an orbital trajectory 
on the $y=\Dy$ plane as it reaches a limit cycle solution in the $\lp x, z\rp$ 
space (Fig.~\ref{fig:d_v_time}d, f), implying that the C$2$ symmetry and time invariance are also
broken.
At $\Ca=0.29$, the system recovers steadiness and concentricity, $\Dy=\Dxz=0$. 
These scenarios suggest the appearance of bifurcating solutions 
by reducing $\Ca$: above a critical value $\Cac\lp \alpha \rp$, the composite system stays concentric, 
corresponding to a stable fixed point solution; it bifurcates 
across $\Cac\lp \alpha \rp$ towards a steady spanwise migration (SM) and/or in-plane orbiting (IPO) motion.

\begin{figure*}[!ht]
{        
    \centering    
    \hspace{3em}\includegraphics[scale = 0.49] 
{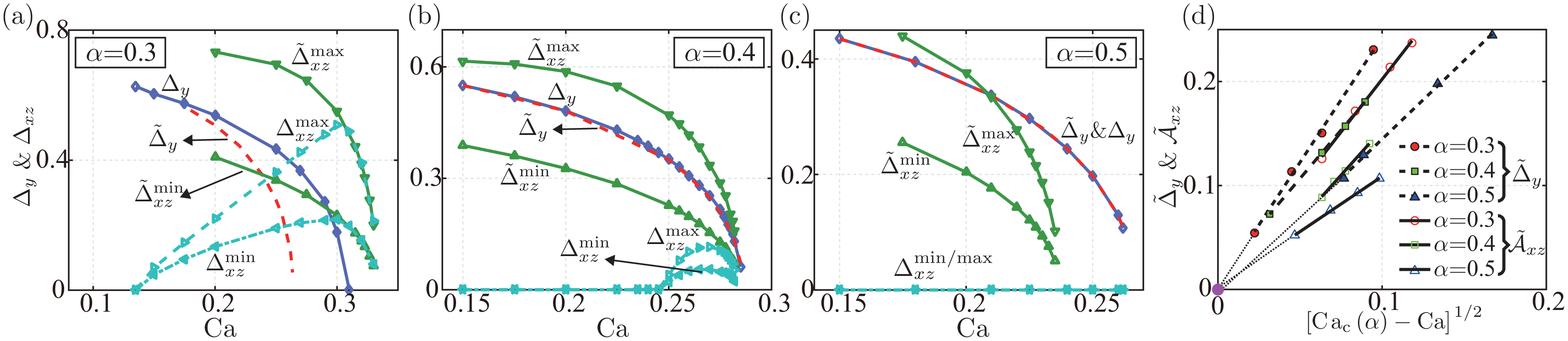}
\caption{Equilibrium displacements $\Dy$, $\Dxzminmax$
versus $\Ca$ for $\alpha=0.3$ (a), $0.4$ (b) and $0.5$ (c); $\tilde{}$ denotes their counterparts of the decoupled simulations. (d):
Linear fitting of $\Dytil$ and $\Axztil=\Dxzmaxtil - \Dxzmintil$ versus $\left[ \Cac \lp \alpha \rp -\Ca \right]^{1/2}$.}
\label{fig:amp_v_ca}
}
\end{figure*}

An investigation spanning the $\lp \Ca,\alpha \rp$ space
further reveals that these two modes, \textit{i.e.} SM and IPO,
appear spontaneously through supercritical pitchfork and Hopf bifurcations respectively.
To study the evolution of the two modes individually,
we perform decoupled simulations with kinematic constraints of
either $\dxz=0$ (pure SM) or $\dy=0$ (pure IPO). Their corresponding 
equilibrium displacements $\Dytil$ and $\Dxzminmaxtil$ 
are shown in Fig.~\ref{fig:amp_v_ca}. 
For all $\alpha$, we observe $\Dytil$ decreases with $\Ca$, becoming zero when $\Ca$
exceeds a critical value $\Cac \lp \alpha \rp$,
so does $\Dxzminmaxtil$. They both vary quadratically in the vicinity of their corresponding
$\Cac \lp \alpha \rp$. This is confirmed by the linear fitting of $\Dytil$ and, $\Axztil = \Dxzmaxtil - \Dxzmintil$ indicating
the oscillating amplitude, 
versus $\left[ \Cac \lp \alpha \rp- \Ca \right]^{1/2}$ (Fig.~\ref{fig:amp_v_ca}d), where $\Cac\lp \alpha \rp$
is obtained simultaneously. The successful fitting passing through the origin 
verifies the emergence of the two bifurcations.
It is worth-pointing that broken reflection symmetry by SM is indeed the signature of pitchfork bifurcation,
so as broken time invariance by IPO of Hopf bifurcation.

Let us return to the constraint-free cases, where the nonlinear interaction of the two modes
results in a more complex dependence of the equilibrium solutions on $\Ca$ and $\alpha$ 
(Fig.~\ref{fig:amp_v_ca}).
For $\alpha=0.3$, the in-plane amplitudes (cyan) reach their maxima around $\Ca = 0.3$,
from where they decrease almost linearly/quadratically with decreasing/increasing $\Ca$.
In their quadratic parts, the coupled (cyan) and decoupled amplitudes (green) overlap in the vicinity 
of their common critical point. In contrast, the spanwise offsets (purple) are larger than those of 
the decoupled cases (red).
For $\alpha=0.4$, the spanwise offsets
coincide precisely with the decoupled counterparts for all $\Ca$, while the
in-plane amplitudes exhibit non-monotonic $\Ca$-dependence as for $\alpha=0.3$
and they are below the decoupled values. For $\alpha=0.5$,
perfect coincidence between the spanwise offsets also holds as in the $\alpha=0.4$ case,
while $\Dxzminmax \equiv 0$ for all $\Ca$, \textit{i.e.}, the Hopf 
bifurcation is inhibited.

\begin{figure}[!ht]
{        
    \centering    
    \hspace{-1em}\includegraphics[scale = 0.54] 
{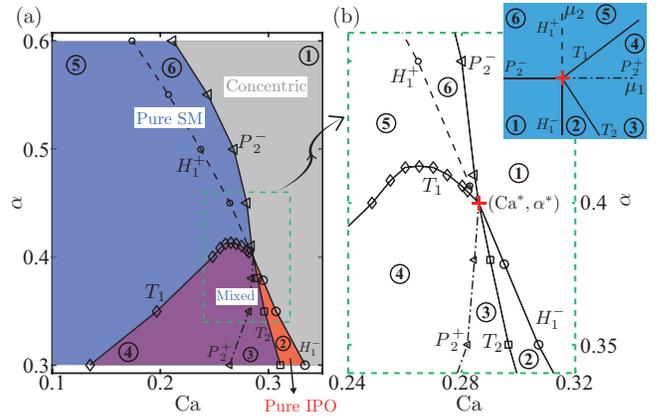}
\caption{(a) Parametric portrait 
 displaying the loci of four solution types in the $\lp \Ca,\alpha \rp$ space. (b) Close-up of (a) in the vicinity of the codimension-two
point $\lp \Caco,\alco \rp$ from where six bifurcations curves originate. 
The inset shows the bifurcation topology in the $\lp \mu_1,\mu_2 \rp$ space.}
\label{fig:phase_map}
}
\end{figure}

The complexity is better unraveled by 
the parametric portrait  quartering the
 $\lp \Ca,\alpha \rp$ parameter space into the following solution types (Fig.~\ref{fig:phase_map}): 'concentric' implying the 
 absence of both modes, 'pure IPO', 'pure SM', and 'mixed' indicating
the coexistence of both modes. A codimension-two point $\lp \Caco,\alco \rp \approx \lp 0.286,0.4 \rp$
is pinpointed  at the intersection of the two marginal curves $H_1^{\pm}$ (circle)
and $P_2^{\pm}$ (triangle) which correspond to
 the Hopf and pitchfork bifurcations, respectively. 
Other branches bifurcating from this point are
$T_1$ separating 'pure SM' and 'mixed', $T_2$ separating 'pure IPO' and 'mixed'. 
Note $^{+}$ and $^{-}$ denote the upper and lower branches of the marginal curves.

We now interpret the bifurcation in the neighborhood of the
 codimension-two point $\lp \Caco, \alco \rp$
based on a normal-form analysis. 
By coupling the amplitude equations of the Hopf and pitchfork
bifurcations, we obtain a normal form similar to that
of the Hopf-Hopf bifurcation in Ref.~\cite{kuznetsov2013elements} where
the amplitudes are independent of phase evolution.
Denoting the square of the in-plane
and spanwise amplitudes by $\rho_1=\lp\Dxzmax-
\Dxzmin \rp^2$ and $\rho_2=\Dy^2$, 
the truncated amplitude system is expressed as

\begin{eqnarray}\label{eq:amp1}
 \dot{\rho}_1 & = & \rho_1 \lp \mu_1 + p_{11} \rho_1 + p_{12} \rho_2 \rp, \nonumber \\
 \dot{\rho}_2 & = & \rho_2 \lp \mu_2 + p_{21} \rho_1 + p_{22} \rho_2 \rp,
\end{eqnarray}
where $\mu_i$ represent the linear growth-rates of the individual modes  and 
$p_{ij}$ the nonlinear coupling coefficients.
Because of the supercritical nature of the two bifurcations, $p_{11}<0$ and $p_{22}<0$. Physically, the amplitudes tend to asymptotic 
values with decreasing $\Ca$ (see Fig.~\ref{fig:amp_v_ca}) owing to
the confinement of droplet. Because $p_{11}p_{22}>0$, 
our problem is in the category of the so-called \textit{simple} cases,
for which we have neglected $\rho_1 \rho^2_2$ and 
$\rho_2 \rho^2_1$ without changing the bifurcation topology~\cite{kuznetsov2013elements}. By introducing 
new phase variables $\xi_1 = -p_{11} \rho_1$ and $\xi_2 = -p_{22} \rho_2$, we obtain

\begin{eqnarray}\label{eq:amp2}
 \dot{\xi}_1 & = & \xi_1 \lp \mu_1 - \xi_1 - \theta \xi_2 \rp, \nonumber \\
  \dot{\xi}_2 & = & \xi_2 \lp \mu_2 - \delta \xi_1 - \xi_2 \rp,
\end{eqnarray}
where $\theta = p_{12}/p_{22}$ and $\delta = p_{21}/p_{11}$.
Applying at leading order the affine transformation

\begin{eqnarray}
 \mu_1 & = & k_1 \lp \Ca -\Caco \rp - \lp \alpha -\alco \rp , \nonumber \\
 \mu_2 & = & C \left [k_2 \lp \Ca -\Caco \rp - \lp \alpha -\alco \rp \right],
\end{eqnarray} 
in the vicinity of $\lp \Caco,\alco \rp$,
we map the parameter space from  $\lp \Ca, \alpha \rp$ to $ \lp \mu_1, \mu_2 \rp$
(inset of Fig.~\ref{fig:phase_map}b),
where $ k_1 \approx -2.3 $ and $k_2 \approx -40$ denote the slope of $H_1$ and $P_2$ curves
at $\lp \Caco,\alco \rp$, with $C \approx 0.043$  derived 
from the growth rate of $\rho_i$. The slopes of $T_1$ and $T_2$ further 
determine
$\theta \approx 0.71 $ and $\delta \approx -0.69$.
 The parametric portrait therefore corresponds to case III described in Ref.~\cite{kuznetsov2013elements}, characterized
by six regions: \rone corresponding to 'concentric'; \rtwo 
to 'pure IPO';  \rthr and \rfou to 'mixed' separated
by $\hsp$; \rfif and \rsix to 'pure SM' separated
by $\hfp$ (see Supplemental material for their phase portraits).

The parametric portrait and normal form both reveal nonlinear mode interactions as a fingerprint of
the present bifurcation. In the absence of SM, IPO appears stably in 
regions \rtwo $\sim$ \rfif but it is suppressed due to the nonlinear interaction with SM, 
as reflected by the phase portraits of \rthr $\sim$ \rfif all including 
an unstable saddle-node equilibrium $\lp \mu_1, 0 \rp$, as well as by the sign of $\theta$. Consequently, pure IPO only survives
 in \rtwo. Besides, without IPO, pure SM is stable in regions \rfou $\sim$ \rsix, while
IPO promotes SM to expand its locus further to \rthr that indeed involves
 a stable equilibrium $\lp \mu_1, \mu_2 \rp$. This promotion results from the
 sign of $\delta$.

 \begin{figure}[!h]
{        
    \centering    
    \hspace{-1em}\includegraphics[scale = 0.35] 
{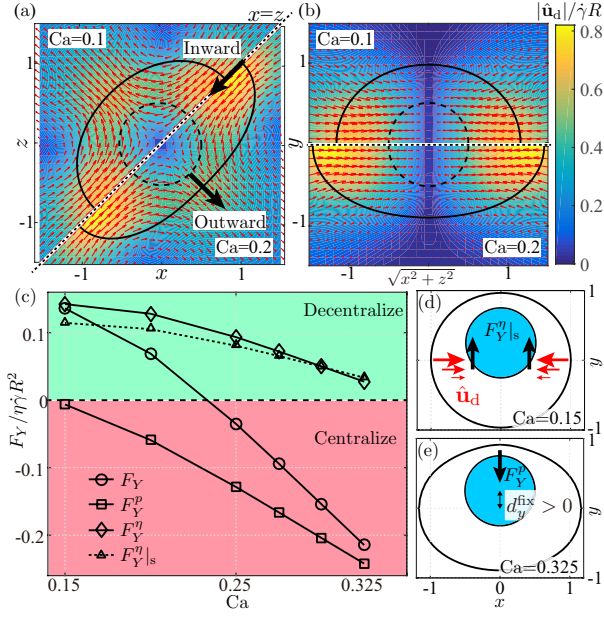}
\caption{Droplet-induced
disturbance flow $\hat{\mathbf{u}}_{\mathrm{d}}$ of concentric systems in the extensional flow,
shown on the $y=0$ (a) and $x=z$ (b) plane,
for $\alpha=0.5$, $\Ca=0.1$ and $0.2$.  
(c)  hydrodynamic force $F_Y\lp \Ca \rp$ exerted on the particle with
a fixed offset $\lp 0, \dy^{\mathrm{fix}}>0, 0 \rp$  inside the droplet under shear; 
$F_Y^{p}$ and $F_Y^{\eta}$ denote the pressure and viscous parts of $F_Y$ respectively, and $F_Y^{\eta}|_{s}$ the shear
force of  $F_Y^{\eta}$. The composite profile is shown on the $z=0$ plane for $\Ca=0.15$ (d) and $0.325$ (e).}
\label{fig:dist_flow}
}
\end{figure}

We next reveal the mechanisms underlying the bifurcations, firstly focusing
on $H_1$ and $P_2$ separately. The shear flow can be decomposed into a rotational and extensional part,
and we found that the former alone does not contribute to the particle's cross-stream motions.
The eccentricity is mostly driven by the extensional part
$\bUinfE \lp \bx \rp= \bE \cdot \bx$ with $E_{xz} = E_{zx}=\dot{\gamma}/2$.
For a system with imposed concentricity in $\bUinfE \lp \bx \rp$, 
Fig.~\ref{fig:dist_flow} displays the droplet-induced
disturbance flows $\udrop$ on $y=0$ (a) and $x=z$ (b) plane, 
which preserve reflection symmetries about each other.
The major/minor axis of the ellipsoid-shaped droplet
lies on $x=\pm z$ plane. The disturbance flow is induced to satisfy
zero normal velocities on the interface. On the shear plane, it approaches/leaves 
the origin along the major/minor axis. 
It resembles the stagnation point flow, where the origin is kinematically unstable. 
This initiates the in-plane motion and indeed the particle moves along 
the minor axis and eventually touches the droplet for any $\Ca$ if we
free its in-plane motion. 
This scenario is altered by the rotational flow, which relocates the particle between
the two axes cyclically. Consequently, it is centralized/decentralized by the
inward/outward flow after every relocation. The inward and outward flows roughly balance at
$\Ca=0.1$; while the former dominates the latter at $\Ca=0.2$, hence overcoming the kinematic instability and 
leading to a concentric preference. 
This might explain the quenching of IPO when $\Ca$ increases across the marginal $H_1$ curve. 
On the $x=z$ plane (Fig.~\ref{fig:dist_flow}b), the flow 
resembles a parallel compressional flow which 
reaches the maximum strength at $y=0$ and weakens in $\pm y$ directions.
When the particle undergoes a spanwise $\dy$ perturbation (say $\dy>0$), 
it experiences the strongest compression
on its lower part and the $y$-gradient of that compressional flow will produces a viscous shear 
force in the spanwise direction (see discussion below) 
that further amplifies this perturbation, 
triggering the pitchfork bifurcation. Note that the flow 
on the $x=-z$ plane may conversely help centralize the particle, yet, it is weaker for any $\Ca>0$. 
Moreover, we conduct simulations fixing a certain spanwise
offset $\dy^{\mathrm{fix}}>0$ with $\dx=\dz=0$ for the shear flow, recording 
the spanwise hydrodynamic forces $F_Y=F_Y^{p} + F_Y^{\eta}$ 
on the particle (Fig.~\ref{fig:dist_flow}c), where $F_Y^{p}$ (resp. $F_Y^{\eta}$) 
represents the pressure (resp. viscous) contribution
which centralizes (resp. decentralizes) the particle.
As shown, the viscous shear force $F_Y^{\eta}|_s$ accounts for the major contribution to $F_Y^{\eta}$,
supporting the above arguments of compression-induced viscous destabilization.
The pressure force becomes stronger with 
$\Ca$ and dominates the viscous part when $\Ca$ exceeds a critical
value. In fact, the droplet with larger $\Ca$ displays a lateral
protrusion accompanying a local curvature increase (Fig.~\ref{fig:dist_flow}e), generating 
a stronger pressure to center the particle. 
This clarifies why SM vanishes when $\Ca$ crosses $P_2$ curve. 
Upon having elucidated $H_1$ and $P_2$ bifurcations individually, we comment
on $T_1$ and $T_2$ which involve mode interactions. 
The trajectory of IPO lies on $y=\Dy$ plane, hence bounded within a circular orbit of 
radius $\lp 1 -\Dy^2\rp^{1/2}-\alpha$ approximately, because the particle simply cannot penetrate the
droplet. The SM mode hence suppresses the IPO mode due to the confinement;
a larger $\Dy$ and/or $\alpha$ naturally shrinks the orbital displacement $\Dxz$ to be zero, 
when entering \rfif across $T_1$. On the contrary, the emergence of $T_2$ reflects the promotive
effect of IPO on SM. Regarding this, we may surmise that when the particle starts orbiting on $y=0$ plane, it
comes closer to the droplet interface; therefore, it suffers a greater compressional flow 
(as indicated by Fig.~\ref{fig:dist_flow}b) which results in stronger destabilizing viscous 
shear forces.

In summary, we have presented in this Letter, hydrodynamic-interaction-meditated
dynamics of a particle inside a droplet in steady shear flow.
We have numerically discovered several equilibrium 
solutions where the composite system exhibits spontaneous symmetry breaking and unsteady dynamics rising through 
supercritical pitchfork and Hopf bifurcations; the particle can
execute spanwise migratory and/or in-plane orbital movement. The bifurcations
are partially attributed to the droplet-induced disturbance flow characterized by 
a kinematically unstable stagnation point. We have performed a normal-form analysis
 to delineate the interplay between bifurcations, revealing
the suppression of the Hopf bifurcation by migration and promotion of the 
pitchfork bifurcation by orbital motion. The interplay can be rationalized by the
geometric confinement and the disturbance flow.

It is worth-pointing that the bifurcation dynamics might not be directly 
generalized
to the two commonly adopted models of cells, capsule and vesicle featured with elastic
membranes. The in-plane elastic stresses developed on the interface might considerably suppress
the interior flow that influences the inclusion dynamics.

We envision that our results might potentially inspire new approaches of 
'hydrodynamic centering' composite systems
like emulsions to obtain a uniform shell in addition to electric centering methods~\cite{bei2008electric,tucker2010polymerization},
or \textit{vice versa},  using hydrodynamic effect to generate 
emulsions with pre-designed nonuniform shell thickness~\cite{hennequin2009synthesizing} 
for programmed release of substances. We hope our study will motivate 
experiments in these directions.
We plan to address in our future work the influences of non-uniform 
shear, geometric features and confinement of the setup, which are all relevant 
for practical applications.

The authors thank Jan Guzowski, J\'{e}r\^{o}me  Hoepffner,  Philippe Meliga, 
Arne Nordmark and Howard A. Stone for
useful discussions. The computer time is provided by the Swiss National 
Supercomputing Centre (CSCS) under project ID s603 and by SNIC (Swedish 
National Infrastructure for Computing). A VR International Postdoc 
Grant from Swedish Research Council '2015-06334' (L.Z.) and an ERC starting grant 'SimCoMiCs 
 280117' (F.G.) are gratefully acknowledged.

\end{document}

%% file: symdef.inc
\def \bUinf {\mathbf{U}^{\infty}}
\def \lp {\left (}
\def \rp {\right )}
\def \bG {\mathbf{G}}
\def \bE {\mathbf{E}}

\def \bx {\mathbf{x}}
\def \Ca {\mathrm{Ca}}
\def \Cac {\Ca_{\mathrm{c}}}
\def \Caco {\Ca_{\ast}}
\def \shr {\dot{\gamma}}
\def \br {\mathbf{d}}

\def \alco {\alpha_{\ast}}

\def \dxz {d_{xz}}
\def \dy {d_{y}}
\def \dx {d_{x}}
\def \dz {d_{z}}

\def \Dy {\Delta_{y}}
\def \Dxz {\Delta_{xz}}

\def \Dxzmax {\Delta^{\mathrm{max}}_{xz}}
\def \Dxzmin {\Delta^{\mathrm{min}}_{xz}}
\def \Dxzminmax {\Delta^{\mathrm{min}/\mathrm{max}}_{xz}}

\def \Dytil {\tilde{\Delta}_{y}}

\def \Dxzmaxtil {\tilde{\Delta}^{\mathrm{max}}_{xz}}
\def \Dxzmintil {\tilde{\Delta}^{\mathrm{min}}_{xz}}
\def \Dxzminmaxtil {\tilde{\Delta}^{\mathrm{min}/\mathrm{max}}_{xz}}

\def \Axztil {\tilde{\mathcal{A}}_{xz}}

\def \bU {\mathbf{U}}
\def \bUinf {\bU^{\infty}}
\def \bUinfE {\bU^{\infty}_{\mathrm{E}}}

\def \hfp {H_1^{+}}

\def \hsp {P_2^{+}}

\def \rone {\circled{$\mathbf{1}$}\;}
\def \rtwo {\circled{$\mathbf{2}$}\;}
\def \rthr {\circled{$\mathbf{3}$}\;}
\def \rfou {\circled{$\mathbf{4}$}\;}
\def \rfif {\circled{$\mathbf{5}$}\;}
\def \rsix {\circled{$\mathbf{6}$}\;}
\def \udrop {\hat{\mathbf{u}}_{\mathrm{d}}}